\documentclass[12pt]{iopart}
\pdfoutput=1
\usepackage{bm}
\usepackage{amssymb}
\usepackage{hyperref}
\usepackage{datetime}
\def\d{{\mathrm{d}}}

\def\O{{\mathcal{O}}}

\makeatletter
\begin{document}

\title[Lorentz invariance and the zero-point stress-energy tensor]{
\centerline{Lorentz invariance and the zero-point stress-energy tensor}
}
\author{Matt~Visser}
\address{School of Mathematics and Statistics,
Victoria University of Wellington; \\
PO Box 600, Wellington 6140, New Zealand.}
\begin{abstract}\\
Some 67 years ago (1951), Wolfgang Pauli noted that the net zero-point energy density could be set to zero by a carefully fine-tuned cancellation between bosons and fermions. In the current article, I will argue in a slightly different direction: the zero-point energy density is only one component of the zero-point stress energy \emph{tensor}, and it is this \emph{tensor} quantity that is in many ways the more fundamental object of interest. I shall demonstrate that \emph{Lorentz invariance} of the  zero-point stress energy tensor implies \emph{finiteness}  of the zero-point stress energy tensor, and vice versa.
Under certain circumstances (in particular, but not limited to, the finite quantum field theories [QFTs]), Pauli's cancellation mechanism will survive the introduction of particle interactions. 
I shall then relate the discussion to beyond standard model [BSM] physics, to~the cosmological constant, and to Sakharov-style induced gravity.


\medskip
\noindent
 arXiv:1610.07264 [gr-qc] 

\medskip
\noindent
D{\sc{ate}}:  24 October 2016; 4 November 2016; 17 November 2016; 24 May 2018\\
\qquad\quad \LaTeX-ed \today 

\medskip
\noindent
P{\sc{ublished}}: Particles {\bf 1} (2018) 10.
\end{abstract}

\maketitle
\tableofcontents
\markboth{Lorentz invariance and the zero-point stress-energy tensor}{}
\clearpage
\section{Introduction}
\def\d{{\mathrm{d}}}

In his ETH lectures of 1951, (transcribed and translated into English in 1971),  Wolfgang Pauli noted that the zero-point energy density could be set to zero by imposing a carefully fine-tuned cancellation between bosons and fermions~\cite{Pauli}. In more modern notation he observed that for relativistic QFTs on a Minkowski background one has:
\begin{equation}
\rho_{zpe} =   \sum_n \left\{  (-1)^{2S_n}  g_n \int {\d^3k\over(2\pi)^3} \, {1\over 2} \, \hbar\omega_n(k) \right\}. 
\label{E:Pauli}
\end{equation}
This integrates the zero-point energy $\pm{1\over 2} \hbar\omega(k)$ over all modes (all three-momenta), counting boson contributions as positive and fermion contributions as negative. The degeneracy factor $g$ includes a spin factor $g = 2S+1$ for massive particles, whereas the spin factor is $g=2$ for massless particles.  The degeneracy factor $g$ also includes an additional factor of $2$ when particle and antiparticle are distinct, and an additional factor of 3 due to colour. (So for example, $g=2$  for the photon, $g=4$ for the electron, and $g=12$ for quarks.) Finally one sums over all particle species indexed by $n$. It is the physical relevance of this sum over the entire particle physics spectrum that is Pauli's key insight. (In related discussion in reference~\cite{Visser:1995} the $(-1)^{2S}$ has been absorbed into the degeneracy factor $g$.) 
Later~on, in the article, we shall directly relate the zero-point energy density $\rho_{zpe}$ to the contentious issue of particle-physics estimates of the cosmological constant (See, for instance, Refs.~\cite{Weinberg:1987, Weinberg:1988, Carroll:1991, Weinberg:1996, Martel:1997, Weinberg:2000a, Weinberg:2000b, Carroll:2000, Peebles:2002, Padmanabhan:2002, Weinberg:2005, Padmanabhan:2007}, but,~for now, let us focus on the the zero-point energy density itself).

Explicitly introducing particle masses one sees
\begin{equation}
\rho_{zpe} =    \sum_n \left\{  (-1)^{2S_n}  g_n\; {1\over 2} \hbar  \int {\d^3k\over(2\pi)^3} \sqrt{m_n^2+k^2} \right\}. 
\end{equation}
Historically, Pauli then imposed a hard momentum cutoff, $k \leq K$, under which the key integral is
\begin{eqnarray}
\label{E:Pauli-rho}
\fl
\int_0^K \d^3k \, \sqrt{m^2+k^2} &=& 
4\pi \int_0^K \d k \; k^2\, \sqrt{m^2+k^2} 
\nonumber\\  \fl&=&
\pi \left\{ K (m^2+K^2)^{3/2} - {1\over2} m^2 K \sqrt{m^2+K^2}  \right.
\nonumber\\ \fl && \qquad\qquad\left. - {1\over2} m^4 \ln\left(K+\sqrt{m^2+K^2}\over m  \right)  \right\}
\nonumber\\ \fl &=&  
\pi \left\{ K^4 + m^2 K^2 +  {m^4\over 8} - {1\over2} m^4\ln(2K/m) \right\} + \O\left(1\over K^2\right).
\end{eqnarray}
(There is an inconsequential and non-propagating typo in Pauli's corresponding formula
in reference~\cite{Pauli}.)
This hard momentum cutoff is certainly adequate in the absence of interactions, which was the situation Pauli was primarily interested in. Subsequently, we shall see various ways of evading the need for any hard momentum cutoff.
Using this integral, Pauli then observed that the total zero-point energy density \emph{vanishes} if and only if we first impose the three polynomial-in-mass conditions
\begin{equation}
\fl\qquad
 \sum_n   (-1)^{2S_n}  g_n = 0;  \qquad  \sum_n  (-1)^{2S_n}  g_n \; m_n^2 =0; 
 \qquad  \sum_n   (-1)^{2S_n}  g_n \; m_n^4=0;
\label{E:polynomial}
\end{equation}
and then supplement this with a fourth logarithmic-in-mass condition
\begin{equation}
 \sum_n   (-1)^{2S_n}  g_n \; m_n^4 \; \ln (m_n^2/\mu^2) = 0.
 \label{E:logarithmic}
\end{equation}

(This fourth logarithmic-in-mass condition is actually independent of the arbitrary parameter $\mu$ due to one already having imposed the third polynomial-in-mass condition.)
This enforced vanishing of the zero-point energy density certainly requires an extremely delicate fine-tuning of the particle physics spectrum.

Let us now modify Pauli's discussion and take it in a rather different direction --- the zero-point energy density is only one component of the zero-point stress-energy tensor; and we shall soon see that this zero-point stress-energy tensor is of more fundamental importance than the zero-point energy density considered in isolation. Specifically, we shall demonstrate that \emph{Lorentz invariance} of the  zero-point stress energy tensor implies \emph{finiteness}  of the zero-point stress energy tensor, and vice versa.

\section{Zero-point stress-energy tensor}
\def\d{{\mathrm{d}}}

The zero-point stress-energy tensor is simply
\begin{equation}
(T_{zpe})^{ab} = \sum_n \left\{  (-1)^{2S_n}  g_n \int {\d^3k\over2\omega_n(k)\; (2\pi)^3} \, \, \hbar \,\, k_n^a \,k_n^b  \right\}. 
\end{equation}
Note that we are now integrating a tensor product of 4-momenta
\begin{equation}
k^a= (\omega(k); k^i)= \left(\sqrt{m^2+k^2}; k^i\right)
\end{equation}
over Lorentz invariant phase space ${\d^3 k/(2\omega)}$. This is certainly Lorentz covariant; making sure it is Lorentz invariant (ie, independent of the inertial frame chosen to do the calculation) is more subtle.  Being more explicit about this
\begin{equation}
\fl
(T_{zpe})^{ab} = \sum_n \left\{  (-1)^{2S_n}  g_n \int {\d^3k\over2\omega_n(k)\; (2\pi)^3} \, \, \hbar \,\,
\left[ \begin{array} {c|c} \omega_n(k)^2 & \omega_n(k) \, k^j\\  \hline \omega_n(k) \, k^i & k^i k^j \end{array} \right]^{ab}  \right\}. 
\end{equation}
Rotational invariance is enough to bring this into the form
\begin{equation}
\fl
(T_{zpe})^{ab} = \sum_n \left\{  (-1)^{2S_n}  g_n \int {\d^3k\over2\omega_n(k)\; (2\pi)^3} \, \, \hbar \,\,
\left[ \begin{array} {c|c} \omega_n(k)^2 & 0\\  \hline 0  & {1\over3} k^2 \,\delta^{ij}\end{array} \right]^{ab}  \right\}. 
\end{equation}
That is, based solely on rotational invariance,
\begin{equation}
\label{E:SET}
(T_{zpe})^{ab} = 
\left[ \begin{array} {c|c} \rho_{zpe} & 0\\  \hline 0  & p_{zpe} \,\delta^{ij}\end{array} \right]. 
\end{equation}
Here the formula for $\rho_{zpe}$ is exactly the same as in Pauli's calculation, equation~(\ref{E:Pauli}), while the zero-point pressure $p_{zpe}$ is seen to be
\begin{equation}
p_{zpe} =    \sum_n \left\{  (-1)^{2S_n}  g_n\; \hbar  \int {\d^3k\over2\sqrt{m_n^2+k^2} \; (2\pi)^3} \; {k^2\over3} \right\}. 
\end{equation}
In references~\cite{Akhmedov:2002, Ossola:2003, Culetu:2004} one encounters rather similar formulae for zero-point energy density and zero-point pressure:
\begin{equation}
\rho_{zpe} =    \pm {\hbar\over2}  \int {\d^3k\over (2\pi)^3} \; \sqrt{m^2+k^2};
\qquad
p_{zpe} =   \pm  {\hbar\over6}  \int {\d^3k \over (2\pi)^3} \; {k^2\over\sqrt{m^2+k^2}}; 
\end{equation}
but without any weighted sum over particle species. See also equation (8) of reference~\cite{Kamenshchik:2016}. 

A somewhat similar formula for the zero-point stress-energy tensor is given in equation (3) of reference~\cite{Mannheim:2011}, and equation (146) of reference~\cite{Mannheim:2016}, but initially without any weighted sum over particle species, and then subsequently regulated by Pauli--Villars ghost terms, (rather than Pauli's weighted sum over physical particle species).
The physical framework considered those articles is somewhat different from that considered in the current article.

\section{Lorentz invariance implies finiteness}

If the zero-point stress energy tensor is to be Lorentz invariant then we must demand $\rho_{zpe}=-p_{zpe}$, or equivalently $\rho_{zpe}+p_{zpe}=0$. But then we have
\begin{equation}
\fl 
\rho_{zpe}+ p_{zpe} =    \sum_n \left\{  (-1)^{2S_n}  g_n\; \hbar  \int {\d^3k\over2\sqrt{m_n^2+k^2} (2\pi)^3} \; \left( \omega_n(k)^2 +{k^2\over3} \right)\right\} = 0.
\end{equation}
That is
\begin{equation}
\fl
\rho_{zpe}+ p_{zpe} =    \sum_n \left\{  (-1)^{2S_n}  g_n\; \hbar  \int {\d^3k\over2\sqrt{m_n^2+k^2} (2\pi)^3} \; 
\left( m_n^2 +{4\over3} k^2\right)\right\} = 0.
\end{equation}
Now observe that
\begin{eqnarray}
\fl
 \int_0^K {\d^3k\over\sqrt{m^2+k^2}} \; \left( m^2 +{4\over3} k^2\right) &=& 
 4\pi \int_0^K {\d k\over\sqrt{m^2+k^2}} \; \left( k^2 m^2 +{4\over3} \, k^4\right)
 \nonumber\\
\fl
 &=& {4\pi\over3} K^3 \sqrt{K^2+m^2}
 \nonumber\\
\fl
 &= & {\pi\over6} \left( 8 K^4 + 4m^2 K^2 - m^4\right) + \O\left(1\over K^2\right).
\end{eqnarray}
Consequently, the zero-point stress-energy tensor is Lorentz invariant if and only if Pauli's three three polynomial-in-mass constraints of equation (\ref{E:polynomial}) are satisfied. (The logarithmic-in-mass constraint of equation~(\ref{E:logarithmic}) need not be satisfied, and in fact the finite value of $\rho_{zpe}$ will be seen to be  proportional to the extent to which this logarithmic-in-mass condition is violated.)
 If these three polynomial-in-mass constraints are satisfied then one has
\begin{equation}
(T_{zpe})^{ab} =  -\rho_{zpe}\; \eta^{ab} = p_{zpe} \;\eta^{ab}.
\end{equation}
Returning to Pauli's analysis for $\rho_{zpe}$, one now sees:
\begin{equation}
\rho_{zpe} = -  p_{zpe} =    {\hbar\over 64\pi^2}  \sum_n   (-1)^{2S_n}  g_n \; m_n^4\;  \ln (m_n^2/\mu^2). 
\label{E:finite}
\end{equation}
Thus Lorentz invariance of the zero-point stress-energy tensor implies finiteness of the zero-point stress-energy tensor. 
Pauli's sum over all particle species is essential to deriving this result.

(For a somewhat related formula see equation (6) of reference~\cite{Mannheim:2011}, or equation (150) of reference~\cite{Mannheim:2016}; but these expressions are regulated by Pauli--Villars ghost terms. 
There is no sum over the physical particle spectrum.)

\section{Finiteness implies Lorentz invariance}

Working in the other direction, if we assume finiteness of the zero-point stress-energy tensor, then in particular $\rho_{zpe}$ must be finite.
Indeed, Pauli's original argument, when applied to \emph{finiteness} of the zero-point energy density, (rather than forcing the zero-point energy density to be \emph{zero}), merely requires that the three polynomial-in-mass constraints of Equation (\ref{E:polynomial}) be enforced. 
Once the third polynomial-in-mass constraint is enforced, the logarithmic cutoff dependence is eliminated.
(See also Ref.~\cite{Alberghi:2008}.)
However, once these three polynomial-in-mass constraints are enforced, the argument of the previous section immediately implies Lorentz invariance.

As a first consistency check, from Lorentz invariance, we see $p_{zpe}=-\rho_{zpe}$, whence the zero point pressure must also be finite.
As a second consistency check, note that the key integral appearing in $p_{zpe}$ is

\begin{eqnarray}
\label{E:Pauli-like-p}
\fl
\int_0^K {\d^3k\over\sqrt{m^2+k^2}} \; k^2 &=& 
 4\pi \int_0^K {\d k\over\sqrt{m^2+k^2}} \; k^4
 \nonumber\\
 \fl
&=& {\pi} \left\{  K^3 \sqrt{K^2+m^2}  - {3\over2} m^2 K \sqrt{K^2+m^2} 
+ {3\over2}m^4 \ln\left(K+\sqrt{m^2+K^2}\over m \right) \right\}
 \nonumber\\
 \fl
 &= & {\pi} \left\{  K^4 - m^2 K^2  -{7\over8} m^4  +{3\over2}m^4\ln\left(2K\over m\right)\right\}
 + \O\left(1\over K^2\right)\!\!.\quad\qquad
\end{eqnarray}

Thus, demanding that $p_{zpe}$ be finite again merely requires that the three polynomial-in-mass constraints of Equation (\ref{E:polynomial}) be enforced. 
(Once the third polynomial-in-mass constraint is enforced the logarithmic cutoff dependence is eliminated.)
That is, the finiteness of both $\rho_{zpe}$ and $p_{zpe}$ are controlled by exactly the same conditions: the three polynomial-in-mass Pauli constraints of Equation~(\ref{E:polynomial}).

Overall, we see that demanding finiteness of the zero-point stress-energy tensor implies Lorentz invariance of the zero-point stress-energy tensor, and we again have the explicit formula of Equation~(\ref{E:finite}).
Pauli's sum over all particle species is again essential to deriving this result.

{Another way of rephrasing these results is as follows: collecting all the divergent (and non-divergent)} terms from Equations (\ref{E:Pauli-rho}) and (\ref{E:Pauli-like-p}), and inserting them into (\ref{E:SET}), we have
\vspace{12pt}
\begin{eqnarray}
(T_{zpe})^{ab} &\propto& 
 \left\{  \sum_n(-1)^{2S_n}  g_n \right\} K^4 \left[\begin{array}{c|c}1&0\\ \hline 0 & {1\over3} \delta^{ij} \end{array} \right]
 \nonumber\\
&&
+
 \left\{ \sum_n (-1)^{2S_n}  g_n m_n^2\right\} K^2 \left[\begin{array}{c|c}  1&0\\ \hline 0 & -{1\over3} \delta^{ij}\end{array} \right]
\nonumber\\
&&
- {1\over2} 
 \left\{  \sum_n (-1)^{2S_n}  g_n m_n^4 \ln(2K/m_n)\right\} \left[\begin{array}{c|c}   1&0\\ \hline 0 & - \delta^{ij}\end{array} \right] 
 \\
&&
 +{1\over 8}
  \left\{  \sum_n (-1)^{2S_n}  g_n m_n^4 \right\} \left[\begin{array}{c|c}   1&0\\ \hline 0 & {7\over3} \delta^{ij}\end{array} \right] 
 \nonumber\\
&&
 + \O(1) \left[\begin{array}{c|c}   1&0\\ \hline 0 & - \delta^{ij}\end{array} \right].\nonumber
\end{eqnarray}

From the above, we again see that finiteness implies Lorentz invariance, and Lorentz invariance implies finiteness.
If we choose to \emph{not} impose finiteness, then the Lorentz-breaking $K^4$ term is a traceless $w=1/3$ cutoff-dependent contribution to the stress energy (qualitatively similar to a ``gas'' of massless light-like particles), while the Lorentz-breaking $K^2$ term would correspond to $w=-1/3$, a ``gas'' just on the verge of violating the strong energy condition (SEC).  The $\ln(K)$ term would correspond to a log-divergent contribution to the cosmological constant. (If one chooses to work only with the trace of the stress-energy tensor then the quadratic $K^4$ divergent term drops out;  see Appendix A. Working with the trace of the stress-energy tensor 
permits one to side-step the most divergent term.)

\section{Evading Pauli's hard momentum cutoff}
In order to side-step the need for Pauli's hard momentum cutoff, let us rewrite the zero-point energy density as
\begin{equation}
\rho_{zpe} =  {1\over 2} \hbar  \int {\d^3k\over(2\pi)^3}   \left\{ \sum_n (-1)^{2S_n}  g_n\;  \sqrt{m_n^2+k^2} \right\}. 
\end{equation}

Here, we perform the sum over particle species first, \emph{before} the momentum integration.  
What are the conditions under which this integral converges?
First, note that there is no IR divergence and that in the UV we have
\begin{equation}
\sqrt{k^2+ m^2} =  k + {m^2\over2k} - {m^4\over8k^3} + \O\left(1\over k^5\right).
\end{equation}

Consequently,
\begin{eqnarray}
\fl
\rho_{zpe} &\propto& \int {\d^3k\over(2\pi)^3} \left\{ k \left( \sum_n  (-1)^{2S_n}  g_n\right) 
+ {1\over2k} \left( \sum_n  (-1)^{2S_n}  g_n m^2 \right) 
- {1\over8 k^3} \left(\sum_n  (-1)^{2S_n}  g_n m^2 \right)
\right.
\nonumber\\
\fl &&\qquad\qquad\left.
+ \O\left(1\over k^5\right) \right\}. 
\end{eqnarray}

This integral converges if and only if the first three polynomial-in-mass Pauli sum rules are satisfied, and one does not need Pauli's hard momentum cutoff to see this. Furthermore, with a little more work, we can evaluate the remaining finite piece. Once we impose the first three polynomial-in-mass Pauli sum rules, we have the identity
\begin{equation}
\rho_{zpe} =  {1\over 2} \hbar  \int {\d^3k\over(2\pi)^3}  
\left\{ \sum_n  (-1)^{2S_n}  g_n\;  \left[ \sqrt{m_n^2+k^2} - k - {m_n^2\over2k} + {m_n^4\over8k^3} \right]\right\}. 
\end{equation}

It is convenient to further split this as follows:
\begin{eqnarray}
\fl
\rho_{zpe} &=&  {1\over 2} \hbar  \int {\d^3k\over(2\pi)^3}  
\sum_n \left\{  (-1)^{2S_n}  g_n\;  
\left[ \sqrt{m_n^2+k^2} - k - {m_n^2\over2k} + {m_n^4\over8k(m_n^2+k^2)} \right]\right\}
\nonumber\\
\fl
&&+
{1\over 2} \hbar  \int {\d^3k\over(2\pi)^3}  
\sum_n \left\{  (-1)^{2S_n}  g_n\;  \left[  {m_n^4\over 8 k^3} - {m_n^4\over8k(m_n^2+k^2)} \right] \right\}.
\end{eqnarray}

Now, an explicit integration yields
\vspace{12pt}
\begin{equation}
\int {\d^3k\over(2\pi)^3}  
 \left[ \sqrt{m_n^2+k^2} - k - {m_n^2\over2k} + {m_n^4\over8k(m_n^2+k^2)}\right]
 \propto m_n^4.
\end{equation}

Thus, by the third Pauli sum rule, the first line vanishes once summed over particle species.
The~remaining term is now
\begin{equation}
\rho_{zpe}  = {1\over 16} \hbar  \int {\d^3k\over(2\pi)^3}  
\sum_n \left\{  (-1)^{2S_n}  g_n m_n^4\;  \left[  {1\over  k^3} - {1\over k(m_n^2+k^2)} \right] \right\}.
\end{equation}

Performing the angular integral, we see
\begin{equation}
\rho_{zpe}  = {1\over 32\pi^2} \hbar  \int {\d k} \,k^2
\sum_n \left\{  (-1)^{2S_n}  g_n m_n^4\;  \left[  {1\over  k^3} - {1\over k(m_n^2+k^2)} \right] \right\}.
\end{equation}

However, introducing the arbitrary but fixed scale $\mu_*$ for convenience, we have
\begin{equation}
\int \d k \, k^2 \left[  {1\over  k^3} - {1\over k(m+k^2)} \right] =  \ln\sqrt{(m^2+k^2)/\mu_*^2} - \ln(k/\mu_*).
\end{equation}

The trailing $\ln(k/\mu_*)$ term vanishes (by the third Pauli sum rule) once one sums over all particle species.
Furthermore, the contribution from the upper limit of integration also vanishes by the third Pauli sum rule.
Finally, we have
\begin{equation}
\rho_{zpe}  = {1\over 32\pi^2} \sum_n \left\{  (-1)^{2S_n}  g_n m_n^4 \ln(m_n/\mu_*)\right\}.
\end{equation}

This is the same result that was previously obtained via Pauli's hard momentum cutoff, but, by rearranging the calculation in this way, we have applied the Pauli constraints directly to the integrand before any integration was carried out, thereby recovering finiteness and even the explicit value of $\rho_{zpe}$ without any need for a momentum cutoff.  Similar arguments apply to $p_{zpe}$. With hindsight, Pauli's hard momentum cutoff is simply a convenience; it is not essential to the argument.  With additional hindsight, this rearrangement of Pauli's calculation can be viewed as a ``physical'' unitarity preserving variant of Pauli--Villars regularization---instead of introducing unitarity violating ``ghost'' fields, one is simply trading off physical bosonic and fermionic contributions to the zero-point energy density against each other.

\section{Other QFT regularizations}
{It is worth pointing out that, since 1951, in place of Pauli's hard momentum cutoff, a number of} other regularization schemes for QFT have been developed.  For instance, Pauli--Villars (ghost field) regularization~\cite{Altschul:2004, Brodsky:1998, Brodsky:1999, Gaillard:1994, Slavnov:1977, Gaillard:1998},  analytic (zeta-function) regularization~\cite{Hawking:1976,BVW, Elizalde:1988, Dittrich:1983}, dimensional regularization~\cite{tHooft:1973, Siegel:1979, Bollini:1972, Breitenlohner:1977, tHooft:1973b}, Schwinger proper-time regularization~\cite{Schwinger:1951a, Schwinger:1951b, Jack:1985, Liao:1994,Bekenstein:1981}, higher-derivative regularizations, (both Lorentz invariant and Lorentz-breaking~\cite{Horava:2009, Sotiriou:2009a, Sotiriou:2009b, Visser:2009, Visser:2009b,Anselmi:2011b, Anselmi:2011a, Anselmi:2010b, Anselmi:2010a, Anselmi:2009b, Anselmi:2009a, Anselmi:2008c, Anselmi:2008b, Anselmi:2008a, Anselmi:2007a, Anselmi:2007b}),  point-splitting regularization~\cite{Christensen:1978, Bunch:1978, Fulling:1981}, and lattice regularization~\cite{Wilson:1974,Kogut:1974}. All of these regularization schemes come at some cost. Two schemes for which the relevant calculations can easily be carried out are zeta-function regularization and dimensional regularization. 

\subsection{Zeta-function regularization}

One considers the zeta-regulated integral~\cite{Hawking:1976,BVW, Elizalde:1988, Dittrich:1983}
\begin{eqnarray}
\rho_{zpe} &\propto& {1\over2} \hbar \mu_* \int {\d^3k\over(2\pi)^3}  \left(m^2+k^2\over\mu_*^2\right)^{1/2 - \epsilon/2}
\propto m^{4-\epsilon}   \int_0^1 x^2 (1+ x^2)^{1/2 - \epsilon/2} \d x
\nonumber\\
&\propto&
m^{4-\epsilon}  {\Gamma(-2+{\epsilon\over2})\over\Gamma(-{1\over2}+{\epsilon\over2})}.
\end{eqnarray}

Here, we set $k=m x$.
The integral converges for $\epsilon>4$. For other values of $\epsilon$, we use properties of the Gamma function to write
\begin{equation}
\rho_{zpe} \propto 
{m^{4-\epsilon} \over \epsilon(2-\epsilon)(4-\epsilon)} {\Gamma(1+{\epsilon\over2})\over\Gamma(-{1\over2}+{\epsilon\over2})}.
\end{equation}

There are manifest poles at $\epsilon=0$,  $\epsilon=2$,  and $\epsilon=4$, and near these poles
\begin{equation}
\fl\qquad
\rho_{zpe} \propto  {-1\over4-\epsilon} + \O(1); \qquad \rho_{zpe} \propto  {-m^2\over2-\epsilon} + \O(1); 
\qquad \rho_{zpe} \propto  {m^4\over\epsilon} + \O(1). 
\end{equation}

If one wishes to analytically continue from the convergent region $\epsilon>4$ to the physical region $\epsilon=0$ using only real values of $\epsilon$, then one should cancel all three poles; doing so moves the abscissa of convergence for the zeta function down to the physical region and keeps $\rho_{zpe}$ real at intermediate steps of the computation. Only if one is willing to formally let $\epsilon$ (and $\rho_{zpe}$) dodge into the complex plane, ignore convergence issues and rely utterly on complex analytic continuation, then would it be sufficient to cancel only the pole at $\epsilon=0$.

The condition to cancel all three poles is that, when summing over all particle species, one should impose the three polynomial-in-mass Pauli conditions; while to cancel only the pole at $\epsilon=0$, one need only impose the third ($m^4$) Pauli condition. This is enough to guarantee finiteness of the zero-point energy density, and the sub-leading $\O(1)$ terms in the expansion around $\epsilon=0$  now give
\begin{equation}
\rho_{zpe} \propto {m^4\over32} - {m^4\ln2\over8} + {m^4\ln m\over8}.
\end{equation}

Summing over particle species and imposing the third polynomial-in-mass Pauli condition, we~have
\begin{equation}
\rho_{zpe} \propto \sum_n (-1)^{2S_n} g_n  \; m_n^4\ln(m/\mu_*).
\end{equation}

This is exactly what is expected based on Pauli's calculations.

\subsection{Dimensional regularization}

Dimensional regularization is based on a formal analytic continuation in the number of dimensions of spacetime~\cite{tHooft:1973, Siegel:1979, Bollini:1972, Breitenlohner:1977, tHooft:1973b}. As such, it is considerably less directly ``physical'' than many other regularization techniques, though computationally, it is typically considered to be the most efficient in terms of evaluating Feynman diagrams.
For the zero-point energy density, the technical details of the calculation for dimensional regularization are formally similar to (but not identical to) those for zeta function regularization. 
One considers the dimensionally regulated integral~\cite{tHooft:1973, Siegel:1979, Bollini:1972, Breitenlohner:1977, tHooft:1973b}
\begin{equation}
\rho_{zpe} \propto {1\over2} \hbar \int {\d^{3-\epsilon}k\over(2\pi)^{3-\epsilon}}  \left(m^2+k^2\right)^{1/2}.
\end{equation}

The integral converges in the UV for $\epsilon>4$. In that UV-convergent range, the integral does not converge in the IR. (This is the same phenomenon that afflicts tadpole diagrams in dimensional regularization.) The standard way of dealing with this in the literature on dimensional regularization is simply to ignore the IR divergences, and, (using $k=mx$) to formally write
\begin{equation}
\rho_{zpe}
\propto m^{4-\epsilon}   \int_0^1 x^{2-\epsilon} (1+ x^2)^{1/2} \d x\propto
m^{4-\epsilon} \; {\Gamma\left(-2+{\epsilon\over2}\right)\Gamma\left({3\over2}+{\epsilon\over2}\right)}.
\end{equation}

The physics reason for ignoring the IR divergences is this: the original physical quantity one is trying to regularize manifestly has no IR divergence---so any IR divergence that arises can only be a physically harmless artefact of the regularization process.
Using properties of the Gamma function, we now write
\begin{equation}
\rho_{zpe} \propto 
{m^{4-\epsilon} \over \epsilon(2-\epsilon)(4-\epsilon)} 
{\Gamma\left(1+{\epsilon\over2}\right) \,\Gamma\left({3\over2}+{\epsilon\over2}\right)}.
\end{equation}

This is quite similar to what we saw for zeta-function regularization---there are again manifest poles at $\epsilon=0$, $\epsilon=2$, and $\epsilon=4$, and near these poles
\begin{equation}
\fl\qquad
\rho_{zpe} \propto  {-1\over4-\epsilon} + \O(1); \qquad \rho_{zpe} \propto  {-m^2\over2-\epsilon} + \O(1); 
\qquad \rho_{zpe} \propto  {m^4\over\epsilon} + \O(1). 
\end{equation}

If one wishes to analytically continue from the UV convergent region $\epsilon>4$ to the physical region $\epsilon=0$ using only real values of the dimension, then one should cancel all three poles. Only if one is willing to formally let the dimension of spacetime (and the zero-point energy density) dodge into the complex plane, and rely utterly on intrinsically complex analytic continuation, would it be sufficient to cancel only the pole at $\epsilon=0$.

The condition to cancel all three poles is again that, when summing over all particle species, one should impose the three polynomial-in-mass Pauli conditions; while to cancel only the pole at $\epsilon=0$, one need only impose the third ($m^4$) Pauli condition. This is enough to guarantee finiteness of the zero-point energy density, and the sub-leading $\O(1)$ terms in the expansion around $\epsilon=0$ again yield
\begin{equation}
\rho_{zpe} \propto {m^4\over32} - {m^4\ln2\over8} + {m^4\ln m\over8}.
\end{equation}

Summing over particle species and imposing the third polynomial-in-mass Pauli condition, we~have
\begin{equation}
\rho_{zpe} \propto \sum_n (-1)^{2S_n} g_n  \; m_n^4\ln(m_n/\mu_*).
\end{equation}

This is again exactly what is expected based on Pauli's calculations.

\section{Summary of the Pauli sum rules}
If Pauli's three polynomial-in-mass constraints hold
\begin{equation}
\fl
 \sum_n   (-1)^{2S_n}  g_n = 0;  \qquad  \sum_n  (-1)^{2S_n}  g_n \; m_n^2 =0; 
 \qquad  \sum_n   (-1)^{2S_n}  g_n \; m_n^4=0;
\label{E:polynomial2}
\end{equation}
then the zero-point energy density and zero-point pressure are finite
\begin{equation}
\rho_{zpe} = -  p_{zpe} =    {\hbar\over 64\pi^2}  \sum_n   (-1)^{2S_n}  g_n \; m_n^4 \; \ln (m_n^2/\mu^2);
\end{equation}
\emph{and} the zero-point stress-energy tensor is Lorentz invariant.  Conversely, Lorentz invariance of the zero-point stress-energy tensor implies both finiteness and Pauli's three polynomial-in-mass constraints. 
While Pauli's original calculation made use of a hard momentum cutoff, we have seen that this can be evaded by 
doing the sum over particle species first, investigating the conditions for convergence, and only then performing the momentum integration.  We have also verified that Pauli's calculation can readily be adapted to zeta-function regularization, and can also be analyzed within the framework of dimensional regularization, ultimately leading to the same physical consequences.

\section{Supersymmetry: neither necessary nor sufficient for Pauli's argument}
It is perhaps worthwhile emphasizing the role that supersymmetry does \emph{not} play in Pauli's argument and the considerations of this article. 
Supersymmetry is not \emph{necessary} in order to set up and understand any of the preceding analysis. Specifically, note that Pauli's 1951 lectures pre-date even the earliest versions of supersymmetry by some 20 years~\cite{Golfand:1971, Volkov:1972, Volkov:1973a, Volkov:1973b, Wess:1973, Wess:1974, West:1976}. Certainly, Pauli's sum rules were  known to some of the originators of supersymmetry, so the sum rules did have a historical input into the foundations of supersymmetry, but they are logically orthogonal thereto. 
While unbroken supersymmetry automatically satisfies all of Pauli's constraints, unbroken supersymmetry is also in violent conflict with empirical reality. 
Conversely, broken supersymmetry, (either spontaneously broken or explicitly broken) need not (and often does not) satisfy the second and third ($m^2$ or $m^4$) Pauli constraints. (The first Pauli constraint, since it just counts bosonic and fermionic degrees of freedom, will generally survive supersymmetry breaking.)

On the other hand, the \emph{finite QFTs} developed in the mid 1980s automatically satisfy all of Pauli's sum rules. While the earliest of the finite QFTs to be found are manifestly supersymmetric~\cite{Mandelstam:1982,Seiberg:1988,  Nilles:1983}, the subsequently developed ``most general'' finite QFTs are manifestly non-supersymmetric, in the sense that they are based on supersymmetric theories that are \emph{softly but explicitly broken}~\cite{Howe:1983, Parkes:1983, Parkes:1984, Kazakov:1995, Piguet:1996, Kobayashi:1997}; the~only part of supersymmetry that survives is the equality between bosonic and fermionic degrees of freedom---the first Pauli sum rule.
Thus, in these general finite QFTs, the supersymmetry is, at best, a~useful book-keeping device. 
The \emph{quasi-finite QFTs} of Ref.~\cite{Parkes:1983} require some explicit clarification: in that specific reference, ``finiteness'' refers only to the scattering amplitudes, and the authors are explicitly excluding the zero-point contributions (the vacuum bubbles) from consideration. 
For that reason, they can avoid imposing the third Pauli sum rule, but the third Pauli sum rule must be reinstated if the vacuum bubbles are to be rendered finite.

Now, if desired, one can of course rewrite the sum over the particle spectrum in the Pauli constraints as a ``supertrace''~\cite{Visser:2002},

\begin{equation}
\sum_{n}   (-1)^{2S_n}  \,g_n  \, X_n =   \mathrm{Str}[ X],
\end{equation}
but this is again merely a book-keeping device, it is not in and of itself an appeal to supersymmetry.
That is: supersymmetry, or lack thereof, is at best logically orthogonal to the questions addressed in this article.

\section{Renormalization group flow of the Pauli sum rules}
Now, consider the effect of interactions, and specifically their impact on the Pauli sum rules via the renormalization group flow of the particle masses. In a completely standard manner, in terms of the renormalization scale $\mu$, let us define the dimensionless $\gamma$ functions as
\begin{equation}
\gamma_n = {\partial\ln m_n\over\partial\ln\mu} = {\mu\over m_n}  {\partial m_n\over\partial \mu}.
\end{equation}

Then, the renormalization group flow for the second and third Pauli sum rules go as
\begin{equation}
\mu {\d\over \d\mu} \left( \sum_n   (-1)^{2S_n}  g_n  \;m_n^2 \right)  
= 2  \left( \sum_n   (-1)^{2S_n}  g_n  \;m_n^2 \; \gamma_n \right);
\end{equation}
\begin{equation}
\mu {\d\over \d\mu} \left( \sum_n   (-1)^{2S_n}  g_n  \;m_n^4 \right)  
= 4  \left( \sum_n   (-1)^{2S_n}  g_n  \;m_n^4 \; \gamma_n \right).
\end{equation}

(Note that the first Pauli sum rule, being proportional to $m^0$, is automatically and trivially preserved under renormalization group flow.)
There are then a number of situations under which the second and third Pauli sum rules are also preserved under the action of the renormalization group~flow.

The finite QFTs developed in the mid 1980s~\cite{Howe:1983, Parkes:1983, Parkes:1984, Kazakov:1995, Piguet:1996, Kobayashi:1997}
 have all of the $\gamma_m=0$, so they automatically provide a quite natural framework for a class of QFTs in which Pauli's sum rules are perturbatively stable against radiative corrections. The \emph{explicit but soft} supersymmetry breaking provides a custodial not-quite-symmetry protecting the Pauli sum rules. Indeed, Pauli's three sum rules are a necessary condition for the known finite QFTs, but they are typically not sufficient. 

A weaker condition is this: even if the full QFT is not finite, as long as mass-renormalization is trivial, then the $\gamma_n$ all vanish (while the $\beta$-functions need not vanish), then this is still enough to guarantee preservation of the Pauli sum rules under renormalization group flow.

An even weaker condition is this: suppose merely that all the $\gamma_m$ are equal, $\gamma_m =\gamma$. Physically this corresponds to \emph{mass ratios} not being renormalized, while an overall mass scale does evolve under the renormalization group flow. Under this condition:
\begin{equation}
\mu {\d\over \d\mu} \left( \sum_n   (-1)^{2S_n}  g_n  \;m_n^2 \right)  
= 2 \gamma \; \left( \sum_n   (-1)^{2S_n}  g_n  \;m_n^2 \right);
\end{equation}
\begin{equation}
\mu {\d\over \d\mu} \left( \sum_n   (-1)^{2S_n}  g_n  \;m_n^4 \right)  
= 4 \gamma \;\left( \sum_n   (-1)^{2S_n}  g_n  \;m_n^4 \right).
\end{equation}

Thus, even this very much milder condition  is still enough to guarantee preservation of the Pauli sum rules under renormalization group flow. If the quantities appearing in the Pauli sum rules start off as zero, they will remain zero.

Finally, let us define what we might call ``Pauli-sum-rule-compatible QFTs'', or more simply ``Pauli-compatible QFTs'', by those QFTs that satisfy the conditions:
\begin{equation}
\fl
\sum_n  \left\{  (-1)^{2S_n}  g_n  \right\}  = 0;
\quad
\sum_n  \left\{  (-1)^{2S_n}  g_n  \;m_n^2 \;\gamma_n\right\}  = 0;
\quad
\sum_n  \left\{  (-1)^{2S_n}  g_n  \;m_n^4 \;\gamma_n\right\} = 0.
\end{equation}

{For this entire class of QFTs, the Pauli sum rules are guaranteed to be preserved under renormalization} group flow.
These constraints are certainly stronger than the Pauli sum rules themselves, and effectively amount to the requirement that the Pauli sum rules should hold not just on-shell, but also for the running particle masses. 
I emphasize that since these conditions are certainly a relaxation of the conditions for the existence of finite QFTs, and because we know that finite QFTs exist~\cite{Howe:1983, Parkes:1983, Parkes:1984, Kazakov:1995, Piguet:1996, Kobayashi:1997}, then we know that  these ``Pauli-compatible QFTs'' certainly exist.

\clearpage
\section{Some implications}

The analysis above impacts on a number of wider issues:
\begin{itemize}
\item Beyond standard model (BSM) physics.
\item Naive estimates of the cosmological constant.
\item Renormalization group running of the cosmological constant.
\item Sakharov-style induced gravity.
\item Graviton contributions to the sum rules.
\end{itemize}

\subsection{Beyond standard model physics}
Let us now take the Pauli sum rules seriously as real physics applied to the real universe. 
Then, to describe reality, we should take the standard model and embed it into one of the ``Pauli-compatible QFTs'' as discussed above. 
The three polynomial-in-mass constraints because they involve the entire particle physics spectrum, certainly impact BSM physics. 
In fact by dividing the spectrum into SM and BSM sectors  we can write the Pauli constraints as:
\begin{equation}
 \sum_{BSM}   (-1)^{2S_n}  g_n = -  \sum_{SM}   (-1)^{2S_n}  g_n;  
 \label{E:bsm1}
\end{equation}
\begin{equation}
 \sum_{BSM}   (-1)^{2S_n}  g_n \;m_n^2 = -  \sum_{SM}   (-1)^{2S_n}  g_n \;m_n^2;  
 \label{E:bsm2}
\end{equation}
\begin{equation}
 \sum_{BSM}   (-1)^{2S_n}  g_n \; m_n^4= -  \sum_{SM}   (-1)^{2S_n}  g_n\; m_n^4.
 \label{E:bsm3}
\end{equation}
That is, enforcing Lorentz invariance of the zero-point on-shell stress-energy tensor, and adopting an overall framework (the ``Pauli-sum-rule-compatible QFTs'') in which these conditions are invariant under renormalization group flow,  makes some  definite predictions for the spectrum of BSM particles, \emph{not least being the fact that there must be BSM particles.}
That is, merely using the very fundamental symmetry principle of Lorentz invariance gives us extremely useful information regarding BSM physics. (The current level of analysis is insufficient to deduce anything more specific regarding the interactions of BSM particles, either  with each other or with the SM sector.)

\subsection{Cosmological constant}

It is commonly asserted that the cosmological constant should be identified with the zero-point energy density, and very naively asserted that it should be estimated by setting $\rho_{cc} = \rho_{zpe} \sim M_{Planck}^4$, with $M_{Planck}$ playing the role of the high-energy cutoff $K$ used at intermediate stages of our argument above. This very naive estimate disagrees with empirical observation by a factor of approximately $10^{123}$ and is famously referred to as the worst prediction in particle physics. 
(See, for instance, various references regarding particle-physics aspects of the cosmological constant~\cite{Weinberg:1987, Weinberg:1988, Carroll:1991, Weinberg:1996, Martel:1997, Weinberg:2000a, Weinberg:2000b, Carroll:2000, Peebles:2002, Padmanabhan:2002, Weinberg:2005, Padmanabhan:2007}.)
However, this is also a dangerously misleading estimate---we know that fermions exist, so there will at the very least be some cancellations in the zero-point energy density. Furthermore, imposing Lorentz invariance has given us a rather definite finite and cutoff-independent estimate for the cosmological constant.

Let us take this analysis a little further. Define two energy scales $\mu_{SM}$ and $\mu_{BSM}$, characteristic of the SM and BSM particle spectra, by setting
\begin{equation}
 \sum_{SM}   (-1)^{2S_n}  g_n\; m_n^4 \; \ln (m_n^2/\mu_{SM}^2) = 0;
\end{equation}
\begin{equation}
 \sum_{BSM}   (-1)^{2S_n}  g_n\; m_n^4 \; \ln (m_n^2/\mu_{BSM}^2) = 0.
\end{equation}
We shall now see that, (at least as far as the cosmological constant is concerned), all of the unknown BSM physics can be summarized by the single parameter $\mu_{BSM}$. 

We have:
\begin{eqnarray}
\fl
\rho_{cc}  
&=& \rho_{zpe} = -  p_{zpe} 
\nonumber\\ \fl 
&=&    {\hbar\over 64\pi^2}  \sum_n   (-1)^{2S_n}  g_n \; m_n^4 \;\ln (m_n^2/\mu^2)
\nonumber\\ \fl
&=& {\hbar\over 64\pi^2}  \sum_n   (-1)^{2S_n}  g_n \; m_n^4 \;\ln (m_n^2/\mu_{BSM}^2)
\nonumber\\ \fl 
&=& {\hbar\over 64\pi^2}  \sum_{SM}   (-1)^{2S_n}  g_n \; m_n^4 \;\ln (m_n^2/\mu_{BSM}^2) +
{\hbar\over 64\pi^2}  \sum_{BSM}   (-1)^{2S_n}  g_n \; m_n^4 \;\ln (m_n^2/\mu_{BSM}^2)
\nonumber\\ \fl
&=&
{\hbar\over 64\pi^2}  \sum_{SM}   (-1)^{2S_n}  g_n \; m_n^4 \;\ln (m_n^2/\mu_{BSM}^2)
\nonumber\\ \fl
&=&
{\hbar\over 64\pi^2}  \sum_{SM}   (-1)^{2S_n}  g_n \; m_n^4 \;\ln (m_n^2/\mu_{SM}^2) 
+{\hbar\over 64\pi^2} \sum_{SM}   (-1)^{2S_n}  g_n \; m_n^4 \;\ln (\mu_{SM}^2/\mu_{BSM}^2) 
\nonumber\\ \fl
&=&
{\hbar\over 64\pi^2} \sum_{SM}   (-1)^{2S_n}  g_n \; m_n^4\; \ln (\mu_{SM}^2/\mu_{BSM}^2).
\end{eqnarray}

That is:
\begin{equation}
\rho_{cc}  = \rho_{zpe} = -  p_{zpe} = - {\hbar\over 64\pi^2} \left\{\sum_{SM}   (-1)^{2S_n}  g_n \; m_n^4 \right\} \ln (\mu_{BSM}^2/\mu_{SM}^2). 
\end{equation}
This is a very clean and elegant result. 
As promised, $\mu_{BSM}$ is the only place that unknown BSM physics now enters into the cosmological constant. 
Without fine tuning of the BSM physics, this might still be astrophysically large, but it will certainly not be $10^{123}$ times too large. 

One can of course always (but perhaps somewhat artificially) tune the cosmological constant to zero (or any empirically supported small quantity)  by introducing a (unitary) variant of the ``Pauli--Villars'' regularization proposal dating back to the 1950s --- merely introduce a sufficient number of non-interacting BSM particles of appropriate mass and statistics. (That is, non-interacting except through gravity, and because you are free to choose the statistics appropriately there is no need to violate unitarity.) This observation reduces the particle-physics ``cosmological constant problem'' to a minor irritation, rather than a major embarrassment.

That Lorentz invariance might help ameliorate the quantitative size of the zero-point energy-density contribution to the cosmological constant has previously been mooted. (See for instance reference~\cite{Akhmedov:2002, Ossola:2003, Culetu:2004, Kamenshchik:2016, Mannheim:2011, Mannheim:2016,  Alberghi:2008, Koksma:2011, Asorey:2012}.) It is the combination of Lorentz invariance with Pauli's sum over all particle species that is central to the current analysis. 

\subsection{Renormalization group flow of the cosmological constant}
Under the renormalization group flow, the cosmological constant naively runs as
\begin{equation}
\fl\qquad
\mu {\d\rho_{zpe}\over\d\mu}  =     {\hbar\over 16\pi^2}  \sum_n   (-1)^{2S_n}  g_n \; m_n^4\;  \ln (m_n^2/\mu_*^2) \;\gamma_n +   {\hbar\over 32\pi^2}  \sum_n   (-1)^{2S_n}  g_n \; m_n^2\;  \gamma_n . 
\label{E:cc-flow}
\end{equation}

However, the second term is zero for all Pauli-compatible QFTs, so more simply
\begin{equation}
\mu {\d\rho_{zpe}\over\d\mu}  =     {\hbar\over 16\pi^2}  \sum_n   (-1)^{2S_n}  g_n \; m_n^4\;  \ln (m_n^2/\mu_*^2) \; \gamma_n.
\label{E:cc-flow2}
\end{equation}
Furthermore, for Pauli-compatible QFTs, this is independent of the arbitrary-but-fixed parameter $\mu_*$.
In~particular, for the finite QFTs, or even for those QFTs with no mass renormalization,  the cosmological constant is likewise unrenormalized. If mass \emph{ratios} are unrenormalized (so all $\gamma_n$ are equal, $\gamma_n = \gamma$), 
then
\begin{equation}
\mu {\d\rho_{zpe}\over\d\mu}  =     4 \gamma \; \rho_{zpe},
\label{E:cc-flow3}
\end{equation}
which has the simple solution
\begin{equation}
\rho_{zpe}(\mu) =     \left[m(\mu)\over m(\mu_0)\right]^4 \; \rho_{zpe}(\mu_0), 
\label{E:cc-flow4}
\end{equation}
implying a simple scaling in line with the overall mass scale. 
However, in the general case, even for Pauli-compatible QFTs, one must deal with Equation (\ref{E:cc-flow2}).

\clearpage
\subsection{Sakharov-style induced gravity}
The preceding analysis is strictly speaking a flat-space Minkowski result, but, due to the locally Euclidean nature of spacetime, it will still govern the dominant short-distance physics in curved spacetime.  There will certainly be sub-leading curvature-dependent terms---which are more easily dealt with by a short-distance asymptotic expansion of the heat kernel in terms of Seeley--DeWitt coefficients. This naturally leads to the concept of Sakharov-like induced gravity~\cite{Sakharov:1967} (see particularly the discussion in Refs.~\cite{Adler:1982,Visser:2002}). The present analysis could easily be modified and extended to further elucidate the induced gravity scenario. 

However, note that some care must be taken to add and subtract only finite regulated physically meaningful quantities, before sending the regulator to infinity. See, for example, Ref.~\cite{Visser:2015}. Over-enthusiastic application of curved space (or even flat space) renormalization techniques can easily eliminate the interesting parts of the physics. 

See, for instance, Ref.~\cite{Elias:2015} for a discussion of some of the potential pitfalls. Note particularly Equations (19) and (23), and how they relate to Pauli's analysis in the flat space limit, and Equation~(26) and how they relate to Sakharov-like induced gravity in curved spacetime.  A somewhat related analysis in terms of a curved spacetime version of the K\"all\'en--Lehmann spectral decomposition, and related spectral sum rules, is given in Ref.~\cite{Kamenshchik:2006}. See also related discussion in~\cite{Gruber:2014}.
(More recently, see Refs.~\cite{Kamenshchik:2018,Ejlli:2017}.)

\subsection{Graviton contributions to the sum rules}

There are at least two ways of dealing with possible graviton contributions to the Pauli sum rules: 
\begin{itemize}
\item  
If one takes the view that gravity is emergent, then one possible variant of this idea is that gravity itself need not be quantized (some variant of the  Sakharov approach) and the question is moot.  This is a respectable minority opinion, but may be somewhat unsatisfying.
\item
If one takes the gravitons of  linearized gravity seriously, then, since they are massless,
    only the zeroth-order-in-mass sum rule matters (massless gravitons do not contribute to the 
    quadratic, quartic, and logarithmic in mass Pauli constraints). The graviton contribution 
    to the zeroth-order-in-mass sum rule can be cancelled by massless sterile ferminons 
    that do not couple to standard model particles. This does not necessarily require 
    supersymmetry (though explicitly broken supersymmetry might prove useful as a 
    book-keeping device.)

\end{itemize}

\section{Conclusions}
\def\implies{\Rightarrow}
The key observation of the current article is the central importance of Lorentz invariance in controlling the finiteness of the zero-point stress-energy tensor: 
\begin{itemize}
\item 
Lorentz invariance $\implies$ the three polynomial-in-mass Pauli constraints $\implies$ finiteness.  
\item
Finiteness $\implies$ the three polynomial-in-mass Pauli constraints $\implies$ Lorentz invariance.
\end{itemize}
This deep and intimate connection between the fundamental physical issues of symmetry and finiteness seems rather oddly to not have previously been developed to the extent that it could.

\ack

This research was supported by the Marsden Fund, 
through a grant administered by the Royal Society of New Zealand.



\addcontentsline{toc}{section}{Appendix A: Trace of the zero-point stress-energy tensor}
\section*{Appendix A: Trace of the zero-point stress-energy tensor}

It is sometimes useful to consider the trace of the zero-point stress-energy tensor~\cite{Visser:1995}
\begin{equation}
\fl
T_{zpe} = -\rho_{zpe}+ 3 p_{zpe} =    \sum_n \left\{  (-1)^{2S_n}  g_n\; \hbar  \int {\d^3k\over2\sqrt{m_n^2+k^2} (2\pi)^3} \; \left( -\omega_n(k)^2 +{k^2} \right)\right\}. 
\end{equation}

Doing this, and (indirectly) imposing Lorentz invariance, gives
\begin{equation}
(T_{zpe})^{ab} = {T_{zpe}\over4} \; \eta^{ab}.
\end{equation}

That is, we could evaluate the zero-point contribution to cosmological constant directly in terms of the~trace:
\begin{equation}
\fl\qquad
T_{zpe} = -\rho_{zpe}+ 3 p_{zpe} =   - \sum_n \left\{  (-1)^{2S_n}  g_n\; \hbar  
\int {\d^3k\over2\sqrt{m_n^2+k^2} (2\pi)^3} \; 
m_n^2\right\}. 
\end{equation}

Now, observe that
\begin{eqnarray}
\fl\qquad
 \int_0^K {\d^3k\over\sqrt{m^2+k^2}} \; m^2 &=& 
 4\pi \int_0^K {\d k\over\sqrt{m^2+k^2}} \; m^2 k^2
 \nonumber\\
\fl\qquad
 &=& \pi \left\{ 2m^2  K \sqrt{m^2+K^2} - 2m^4\ln\left(K+\sqrt{m^2+K^2}\over m\right) \right\} 
\\
\fl\qquad
 &= & {\pi} \left\{ 2 m^2 K^2 + m^4 - 2 m^4 \ln\left(2K\over m\right) \right\} + \O\left(1\over K^2\right). \nonumber
\end{eqnarray}

The only minor oddity here is that there is no $K^4$ term. (It is not all that odd an oddity: the explicit presence of the factor of $m^2$  in the integral above guarantees that no $K^4$ term can possibly arise. Taking the trace is a useful trick for evading the biggest potentially divergent contribution, the quartic term. See related discussion in Ref.~\cite{Visser:1995}.) 


Thus:
\begin{itemize}
\item Finiteness of the trace $T_{zpe}=-\rho_{zpe}+ 3 p_{zpe}$ requires only the second and the third of the polynomial-in-mass constraint conditions of Equation~(\ref{E:polynomial}). If we impose these two constraints,~then
\begin{equation}
\fl\qquad
T_{zpe} = - \rho_{zpe}+ 3 p_{zpe} = -{1\over16\pi^2}   \sum_n \left\{  (-1)^{2S_n}  g_n\; \hbar \, m_n^4\,  \ln( m_n^2/\mu_*^2)  \right\}.
\end{equation}
\item Vanishing of the trace $T_{zpe}=-\rho_{zpe}+ 3 p_{zpe}$ requires only the second and the third of the polynomial-in-mass constraint conditions of Equation~(\ref{E:polynomial}), plus the logarithmic-in-mass constraint of Equation (\ref{E:logarithmic}). 
\item {Unfortunately, controlling the trace in this manner is not enough to render the zero-point stress-energy} tensor Lorentz invariant; which requires all three of the polynomial-in-mass constraints as discussed above. (See a somewhat related discussion in~\cite{Visser:1995}.)
\end{itemize}

\section*{References}


\begin{thebibliography}{999}

\bibitem{Pauli}
Wolfgang Pauli, \emph{Pauli Lectures on Physics: Vol 6, Selected Topics in Field Quantization},\\
MIT Press, 1971 (editor C.P. Enz).\\
(Translation of ``Feldquantisierung'' 1950--1951; see especially page 33 of the English translation.)


  \bibitem{Visser:1995}
  M.~Visser,
  ``Lorentzian wormholes: From Einstein to Hawking'', \\
  (AIP Press, now Springer--Verlag, 1995).
  See especially pages 82--84. 




\bibitem{Weinberg:1987}
Weinberg, S. Anthropic Bound on the Cosmological Constant. \emph{Phys.\ Rev.\ Lett.}\ {\bf 1987}, \emph{59}, 2607, doi:10.1103/PhysRevLett.59.2607.
\bibitem{Weinberg:1988}
Weinberg, S. The Cosmological Constant Problem. \emph{Rev.\ Mod.\ Phys.}\ {\bf 1989}, \emph{61}, 1, doi:10.1103/RevModPhys.61.1.


\bibitem{Carroll:1991}
Carroll, S.M.; Press, W.H.; Turner, E.L. The Cosmological constant. \emph{Ann.\ Rev.\ Astron.\ Astrophys.}\ {\bf 1992}, \emph{30}, 499--542, doi:10.1146/annurev.aa.30.090192.002435.
\bibitem{Weinberg:1996}
Weinberg, S. Theories of the cosmological constant. In Proceedings of the Critical Dialogues in Cosmology, Princeton, NJ, USA, 24--27 June 1996; pp. 195--203.

\bibitem{Martel:1997}
Martel, H.; Shapiro, P.R.; Weinberg, S. Likely values of the cosmological constant. \emph{Astrophys.\ J.}\ {\bf 1998}, \emph{492}, 29, doi:10.1086/305016.

\bibitem{Weinberg:2000a}
Weinberg, S. A priori probability distribution of the cosmological constant. \emph{Phys.\ Rev.\ D} {\bf 2000}, \emph{61}, 103505, doi:10.1103/PhysRevD.61.103505.
\bibitem{Weinberg:2000b}
Weinberg, S. The Cosmological constant problems. \emph{arXiv} \textbf{2000}, arXiv:astro-ph/0005265.
\bibitem{Carroll:2000}
Carroll, S.M. The Cosmological constant. \emph{Living Rev.\ Relativ.}\ {\bf 2001}, \emph{4}, 1, doi:10.12942/lrr-2001-1.

\bibitem{Peebles:2002}
Peebles, P.J.E.; Ratra, B. The Cosmological constant and dark energy. \emph{Rev.\ Mod.\ Phys.}\ {\bf 2003}, \emph{75}, 559, doi:10.1103/RevModPhys.75.559.

\bibitem{Padmanabhan:2002}
Padmanabhan, T. Cosmological constant: The Weight of the vacuum. \emph{Phys.\ Rep.}\ {\bf 2003}, \emph{380}, 235--320, doi:10.1016/S0370-1573(03)00120-0.

\bibitem{Weinberg:2005}
Weinberg, S. Einstein's mistakes. \emph{Phys.\ Today} {\bf 2005}, \emph{58}, 31, doi:10.1063/1.2155755.
\bibitem{Padmanabhan:2007}
Padmanabhan, T. Dark energy and gravity. \emph{Gen.\ Relativ.\ Gravit.}\ {\bf 2008}, \emph{40}, 529--564, doi:10.1007/s10714-007-0555-7.



 \bibitem{Akhmedov:2002}
  E.~K.~Akhmedov,
  ``Vacuum energy and relativistic invariance'',
  hep-th/0204048.
  
  \bibitem{Ossola:2003}
  G.~Ossola and A.~Sirlin,\\
  \leftline{``Considerations concerning the contributions of fundamental particles to the vacuum energy density'',}\\
  Eur.\ Phys.\ J.\ C {\bf 31} (2003) 165
  doi:10.1140/epjc/s2003-01337-7
  [hep-ph/0305050].
  
  \bibitem{Culetu:2004}
  H.~Culetu,
  ``The zero point energy and gravitation'',
  hep-th/0410133.\\
  See especially equations (3.1)--(3.2), (3.5)--(3.6), and (3.12). 
  
   \bibitem{Kamenshchik:2016}
  A.~Y.~Kamenshchik, A.~A.~Starobinsky, A.~Tronconi, G.~P.~Vacca and G.~Venturi,\\
  ``Vacuum energy, Standard Model physics and the $750\; \rm{GeV}$ Diphoton Excess at the LHC,''\\
  arXiv:1604.02371 [hep-ph].

 
 
\bibitem{Mannheim:2011}
 P.~D.~Mannheim,\\
  ``Intrinsically quantum-mechanical gravity and the cosmological constant problem'',\\
  Mod.\ Phys.\ Lett.\ A {\bf 26} (2011) 2375
  doi:10.1142/S0217732311036875
  [arXiv:1005.5108 [hep-th]].
  See especially equation (6).
  
  \bibitem{Mannheim:2016}
  P.~D.~Mannheim,\\
  \leftline{``Mass generation, the cosmological constant problem, conformal symmetry, and the Higgs boson'',}\\
  arXiv:1610.08907 [hep-ph].
   See especially equation (150).
  
 
  
  \bibitem{Alberghi:2008}
  G.~L.~Alberghi, A.~Y.~Kamenshchik, A.~Tronconi, G.~P.~Vacca and G.~Venturi,\\
  ``Vacuum energy, cosmological constant and standard model physics,''\\
  JETP Lett.\  {\bf 88} (2008) 705.
  doi:10.1134/S002136400823001X
   
 

\bibitem{Altschul:2004}
Altschul, B. Gauge invariance and the Pauli-Villars regulator in Lorentz- and CPT-violating electrodynamics. \emph{Phys.\ Rev.\ D} {\bf 2004}, \emph{70}, 101701, doi:10.1103/PhysRevD.70.101701.
\bibitem{Brodsky:1998}
Brodsky, S.J.; Hiller, J.R.; McCartor, G. Pauli-Villars as a nonperturbative ultraviolet regulator in discretized light cone quantization. \emph{Phys.\ Rev.\ D} {\bf 1998}, \emph{58}, 025005, doi:10.1103/PhysRevD.58.025005.
\bibitem{Brodsky:1999}
Brodsky, S.J.; Hiller, J.R.; McCartor, G. Application of Pauli-Villars regularization and discretized light cone quantization to a (3+1)-dimensional model. \emph{Phys.\ Rev.\ D} {\bf 1999}, \emph{60}, 054506, doi:10.1103/PhysRevD.60.054506.
\bibitem{Gaillard:1994}
Gaillard, M.K. Pauli-Villars regularization of supergravity coupled to chiral and Yang-Mills matter. \emph{Phys.~Lett.~B} {\bf 1995}, 342, 125--131, doi:10.1016/0370-2693(94)01341-9.
\bibitem{Slavnov:1977}
Slavnov, A.A. The Pauli-Villars Regularization for Nonabelian Gauge Theories. \emph{Theor. Math. Phys.}\ {\bf 1977}, \emph{33}, 977--981, doi:10.1007/BF01036595.
\bibitem{Gaillard:1998}
Gaillard, M.K. One loop Pauli-Villars regularization of supergravity 1. Canonical gauge kinetic energy. \emph{Phys.~Rev.~D} {\bf 1998}, \emph{58}, 105027, doi:10.1103/PhysRevD.58.105027.


\bibitem{Hawking:1976}
Hawking, S.W. Zeta Function Regularization of Path Integrals in Curved Space-Time. \emph{Commun.\ Math.\ Phys.}\ {\bf 1977}, \emph{55}, 133--148, doi:10.1007/BF01626516.
\bibitem{BVW}
Blau, S.; Visser, M.; Wipf, A. Zeta Functions and the Casimir Energy. \emph{Nucl.\ Phys.\ B} {\bf 1988}, \emph{310}, 163, doi:10.1016/0550-3213(88)90059-4.
\bibitem{Elizalde:1988}
Elizalde, E.; Romeo, A. Expressions for the zeta Function Regularized Casimir Energy. \emph{J.\ Math.\ Phys.}\ {\bf 1989}, 30, 1133--1139, doi:10.1063/1.528332.
\bibitem{Dittrich:1983}
Dittrich, W.; Reuter, M. Effective {QCD} Lagrangian With Zeta Function Regularization. \emph{Phys.\ Lett. B}\ {\bf 1983}, \emph{128}, 321--326, doi:10.1016/0370-2693(83)90268-X.




\bibitem{tHooft:1973}
Hooft, G. Dimensional regularization and the renormalization group. \emph{Nucl.\ Phys.\ B} {\bf 1973}, \emph{61}, 455--468, doi:10.1016/0550-3213(73)90376-3.
\bibitem{Siegel:1979}
Siegel, W. Supersymmetric Dimensional Regularization via Dimensional Reduction. \emph{Phys.\ Lett. B}\ {\bf 1979}, \emph{84}, 193--196, doi:10.1016/0370-2693(79)90282-X.
\bibitem{Bollini:1972}
Bollini, C.G.; Giambiagi, J.J. Dimensional Renormalization: The Number of Dimensions as a Regularizing Parameter. \emph{Nuovo Cim.\ B} {\bf 1972}, \emph{12}, 20--26, doi:10.1007/BF02895558.
\bibitem{Breitenlohner:1977}
Breitenlohner, P.; Maison, D. Dimensional Renormalization and the Action Principle. \emph{Commun.\ Math.\ Phys.}\ {\bf 1977}, \emph{52}, 11--38, doi:10.1007/BF01609069.
\bibitem{tHooft:1973b}
Hooft, G. An algorithm for the poles at dimension four in the dimensional regularization procedure. \emph{Nucl.~Phys.\ B} {\bf 1973}, \emph{62}, 444--460, doi:10.1016/0550-3213(73)90263-0.
\bibitem{Schwinger:1951a}
Schwinger, J.S. On gauge invariance and vacuum polarization. \emph{Phys.\ Rev.}\ {\bf 1951}, \emph{82}, 664, doi:10.1103/PhysRev.82.664.
\bibitem{Schwinger:1951b}
Schwinger, J.S. The Theory of quantized fields. I. \emph{Phys.\ Rev.}\ {\bf 1951}, \emph{82},  914, doi:10.1103/PhysRev.82.914. 
\bibitem{Jack:1985}
Jack, I.; Parker, L. Proof of Summed Form of Proper Time Expansion for Propagator in Curved Space-time. \emph{Phys.\ Rev.\ D} {\bf 1985}, 31, 2439, doi:10.1103/PhysRevD.31.2439.
\bibitem{Liao:1994}
Liao, S.B. On connection between momentum cutoff and the proper time regularizations. \emph{Phys.~Rev.~D} {\bf 1996}, \emph{53}, 2020, doi:10.1103/PhysRevD.53.2020.
\bibitem{Bekenstein:1981}
Bekenstein, J.D.; Parker, L. Path Integral Evaluation of Feynman Propagator in Curved Space-time. \emph{Phys.~Rev.~D} {\bf 1981}, \emph{23}, 2850, doi:10.1103/PhysRevD.23.2850.




\bibitem{Horava:2009}
{Horava, P. Quantum Gravity at a Lifshitz Point. \emph{Phys.\ Rev.\ D} {\bf 2009}, \emph{79}, 084008, doi:10.1103/PhysRevD.79.084008.}
\bibitem{Sotiriou:2009a}
Sotiriou, T.P.; Visser, M.; Weinfurtner, S. Phenomenologically viable Lorentz-violating quantum gravity. \emph{Phys.~Rev.~Lett.}\ {\bf 2009}, \emph{102}, 251601, doi:10.1103/PhysRevLett.102.251601.

\bibitem{Sotiriou:2009b}
Sotiriou, T.P.; Visser, M.; Weinfurtner, S. Quantum gravity without Lorentz invariance. \emph{JHEP} {\bf 2009}, \emph{0910}, 033, doi:10.1088/1126-6708/2009/10/033.
\bibitem{Visser:2009}
Visser, M. Lorentz symmetry breaking as a quantum field theory regulator. \emph{Phys.\ Rev.\ D} {\bf 2009}, \emph{80}, 025011, doi:10.1103/PhysRevD.80.025011.
\bibitem{Visser:2009b}
Visser, M. Power-counting renormalizability of generalized Horava gravity. \emph{arXiv} \textbf{2009}, arXiv:0912.4757~[hep-th].

\bibitem{Anselmi:2011b}
Anselmi, D. Renormalization and Lorentz symmetry violation. {In Proceedings of the Workshop on Continuum and Lattice Approaches to Quantum Gravity, Brighton, UK, 17--19 September 2008}; Proceedings of Science; doi:10.22323/1.079.0010.

\bibitem{Anselmi:2011a}
Anselmi, D. Standard Model and High Energy Lorentz Violation. In \emph{{Strong Coupling Gauge Theories in LHC Era}}; {World Scientific, Singapore}, 2011; pp. 357--363, doi:10.1142/9789814329521\_0038.
\bibitem{Anselmi:2010b}
Anselmi, D.; Ciuffoli, E. Renormalization of High-Energy Lorentz Violating Four Fermion Models. \emph{Phys.~Rev.~D} {\bf 2010}, \emph{81}, 085043, doi:10.1103/PhysRevD.81.085043.

\bibitem{Anselmi:2010a}
{Anselmi, D. Standard Model and High-energy Lorentz Violation}. In Proceedings of the Workshop in Honor of Toshihide Maskawa's 70th Birthday and 35th Anniversary of Dynamical Symmetry Breaking in SCGT, Nagoya University, Nagoya, Japan, 8 – 11 December 2009; doi:10.1142/9789814327688\_0041.

\bibitem{Anselmi:2009b}
Anselmi, D.; Taiuti, M. Renormalization of High-Energy Lorentz Violating QED. \emph{Phys.\ Rev.\ D} {\bf 2010}, \emph{81}, 085042, doi:10.1103/PhysRevD.81.085042.

\bibitem{Anselmi:2009a}
Anselmi, D. Standard Model without Elementary Scalars and High Energy Lorentz Violation. \emph{Eur.\ Phys.\ J.\ C} {\bf 2010}, \emph{65}, 523--536, doi:10.1140/epjc/s10052-009-1211-z.

\bibitem{Anselmi:2008c}
Anselmi, D. Weighted power counting, neutrino masses and Lorentz violating extensions of the Standard Model. \emph{Phys.\ Rev.\ D} {\bf 2009}, \emph{79}, 025017, doi:10.1103/PhysRevD.79.025017.

\bibitem{Anselmi:2008b}
Anselmi, D. Weighted power counting and Lorentz violating gauge theories. II. Classification. \emph{Ann. Phys.}\ {\bf 2009}, \emph{324}, 1058--1077, doi:10.1016/j.aop.2008.12.007.

\bibitem{Anselmi:2008a}
Anselmi, D. Weighted power counting and Lorentz violating gauge theories. I. General properties. \emph{Ann.~Phys.}\ {\bf 2009}, \emph{324}, 874--896, doi:10.1016/j.aop.2008.12.005.

\bibitem{Anselmi:2007a}
{Anselmi, D. Renormalization of Lorentz violating theories}. 
Proceedings of the Fourth Meeting on CPT and Lorentz Symmetry
Bloomington, USA, 8--11 August 2007; pp. 219-223.
(World Scientific, Singapore, 2008). 
doi:10.1142/9789812779519\_0032.

\bibitem{Anselmi:2007b}
Anselmi, D.; Halat, M. Renormalization of Lorentz violating theories. \emph{Phys.\ Rev.\ D} {\bf 2007}, \emph{76}, 125011, doi:10.1103/PhysRevD.76.125011.


\bibitem{Christensen:1978}
Christensen, S.M. Regularization, Renormalization, and Covariant Geodesic Point Separation. \emph{Phys.\ Rev.\ D} {\bf 1978}, \emph{17}, 946, doi:10.1103/PhysRevD.17.946.
\bibitem{Bunch:1978}
Bunch, T.S.; Davies, P.C.W. Quantum Field Theory in de Sitter Space: Renormalization by Point Splitting. \emph{Proc.\ R.\ Soc.\ Lond.\ A} {\bf 1978}, \emph{360}, 117--134, doi:10.1098/rspa.1978.0060.
\bibitem{Fulling:1981}
Fulling, S.A.; Narcowich, F.J.; Wald, R.M. Singularity Structure of the Two Point Function in Quantum Field Theory in Curved Space-time. {II}. \emph{Ann. Phys.}\ {\bf 1981}, \emph{136}, 243--272, doi:10.1016/0003-4916(81)90098-1.


\bibitem{Wilson:1974}
Wilson, K.G. Confinement of Quarks. \emph{Phys.\ Rev.\ D} {\bf 1974}, \emph{10}, 2445, doi:10.1103/PhysRevD.10.2445.
\bibitem{Kogut:1974}
Kogut, J.B.; Susskind, L. Hamiltonian Formulation of Wilson's Lattice Gauge Theories. \emph{Phys.\ Rev.\ D} {\bf 1975}, \emph{11}, 395, doi:10.1103/PhysRevD.11.395.




\bibitem{Golfand:1971}
Golfand, Y.A.; Likhtman, E.P. Extension of the algebra of Poincare group generators and violation of $P$ invariance. \emph{JETP Lett.}\ {\bf 1971}, \emph{13}, {323--326}.
\bibitem{Volkov:1972}
Volkov, D.V.; Akulov, V.P. Possible universal neutrino interaction. \emph{JETP Lett.}\ {\bf 1972}, \emph{16}, {438--440}.
\bibitem{Volkov:1973a}
Volkov, D.V.; Akulov, V.P. Is the neutrino a Goldstone particle? \emph{Phys.\ Lett.}\ {\bf 1973}, \emph{46B}, 109--110, doi:10.1016/0370-2693(73)90490-5.
\bibitem{Volkov:1973b}
Volkov, D.V.; Soroka, V.A. Higgs effect for Goldstone particles with spin 1/2. \emph{JETP Lett.}\ {\bf 1973}, \emph{18}, {312--314}.

\bibitem{Wess:1973}
Wess, J.; Zumino, B. A Lagrangian model invariant under supergauge transformations. \emph{Phys.\ Lett. B}\ {\bf 1974}, \emph{49}, 52--54, doi:10.1016/0370-2693(74)90578-4.
\bibitem{Wess:1974}
Wess, J.; Zumino, B. Supergauge invariant extension of quantum electrodynamics. \emph{Nucl.\ Phys.\ B} {\bf 1974}, \emph{78}, 1--13, doi:10.1016/0550-3213(74)90112-6.
\bibitem{West:1976}
West, P.C. The Supersymmetric Effective Potential. \emph{Nucl.\ Phys.\ B} {\bf 1976}, \emph{106}, 219, doi:10.1016/0550-3213(76)90378-3.

\bibitem{Mandelstam:1982}
Mandelstam, S. Light Cone Superspace and the Ultraviolet Finiteness of the $n=4$ Model. \emph{Nucl.\ Phys.\ B} {\bf 1983}, \emph{213}, 149--168, doi:10.1016/0550-3213(83)90179-7.
\bibitem{Seiberg:1988}
Seiberg, N. Supersymmetry and nonperturbative beta functions. \emph{Phys.\ Lett.\ B} {\bf 1988}, \emph{206}, 75--80, 
doi:10.1016/0370-2693(88)91265-8.
\bibitem{Nilles:1983}
Nilles, H.P. Supersymmetry, Supergravity and Particle Physics. \emph{Phys.\ Rept.}\ {\bf 1984}, \emph{110}, 1--162, doi:10.1016/0370-1573(84)90008-5.



\bibitem{Howe:1983}
Howe, P.S.; Stelle, K.S.; West, P.C. A Class of Finite Four-Dimensional Supersymmetric Field Theories. \emph{Phys.~Lett.}\ {\bf 1983}, \emph{124}, 55--58, doi:10.1016/0370-2693(83)91402-8.


\bibitem{Parkes:1983}
Parkes, A.; West, P.C. Explicit supersymmetry breaking can preserve finiteness in rigid $n=2$ supersymmetric theories. \emph{Phys.\ Lett. B}\ {\bf 1983}, \emph{127}, 353--359, doi:10.1016/0370-2693(83)91016-X.
\bibitem{Parkes:1984}
Parkes, A.; West, P.C. Finiteness in rigid supersymmetric theories. \emph{Phys.\ Lett. B}\ {\bf 1984}, \emph{138}, 99--104, doi:10.1016/0370-2693(84)91881-1.
\bibitem{Kazakov:1995}
Kazakov, D.I.; Kalmykov, M.Y.; Kondrashuk, I.N.; Gladyshev, A.V. Softly broken finite supersymmetric grand unified theory. \emph{Nucl.\ Phys.\ B} {\bf 1996}, \emph{471}, 389--408, doi:10.1016/0550-3213(96)00180-0.
\bibitem{Piguet:1996}
Piguet, O. Supersymmetry, ultraviolet finiteness and grand unification. \emph{arXiv} \textbf{1996}, arXiv:hep-th/9606045.
\bibitem{Kobayashi:1997}
Kobayashi, T.; Kubo, J.; Mondragon, M.; Zoupanos, G. Constraints on finite soft supersymmetry breaking terms. \emph{Nucl.\ Phys.\ B} {\bf 1998}, \emph{511}, 45--68, doi:10.1016/S0550-3213(97)00765-7.

\bibitem{Visser:2002}
Visser, M. Sakharov's induced gravity: A modern perspective. \emph{Mod.\ Phys.\ Lett.\ A} {\bf 2002}, \emph{17}, 977--991, doi:10.1142/S0217732302006886.
\bibitem{Koksma:2011}
Koksma, J.F.; Prokopec, T. The cosmological constant and Lorentz invariance of the vacuum state. \emph{arXiv} \textbf{2011}, arXiv:1105.6296 [gr-qc].
\bibitem{Asorey:2012}
Asorey, M.; Lavrov, P.M.; Ribeiro, B.J.; Shapiro, I.L. Vacuum stress-tensor in SSB theories. \emph{Phys.\ Rev.\ D} {\bf 2012}, \emph{85}, 104001, doi:10.1103/PhysRevD.85.104001.






\bibitem{Sakharov:1967}
Sakharov, A.D. Vacuum quantum fluctuations in curved space and the theory of gravitation. \emph{Sov.\ Phys.\ Dokl.}\ {\bf 1968}, \emph{12}, {1040--1041}.
\bibitem{Adler:1982}
Adler, S.L. Einstein Gravity as a Symmetry-Breaking Effect in Quantum Field Theory. \emph{Rev.\ Mod.\ Phys.}\ {\bf 1982}, \emph{54}, 729, doi:10.1103/RevModPhys.54.729. 

\bibitem{Visser:2015}
Visser, M. Why are Casimir energy differences so often finite? \emph{arXiv} \textbf{2016}, arXiv:1601.01374 [quant-ph].
\bibitem{Elias:2015}
El\'ias, M.; Mazzitelli, F.D. Ultraviolet cutoffs for quantum fields in cosmological spacetimes. \emph{Phys.\ Rev.\ D} {\bf 2015}, \emph{91}, 124051, doi:10.1103/PhysRevD.91.124051.


\bibitem{Kamenshchik:2006}
Kamenshchik, A.Y.; Tronconi, A.; Vacca, G.P.; Venturi, G. Vacuum energy and spectral function sum rules. \emph{Phys.\ Rev.\ D} {\bf 2007}, \emph{75}, 083514, doi:10.1103/PhysRevD.75.083514.
\bibitem{Gruber:2014}
Gruber, C.; Kleinert, H. Observed cosmological re-expansion in minimal QFT with Bose and Fermi fields. \emph{Astropart.\ Phys.}\ {\bf 2014}, \emph{61}, 72--81, doi:10.1016/j.astropartphys.2014.06.012.

\bibitem{Kamenshchik:2018}
Kamenshchik, A.Y.; Starobinsky, A.A.; Tronconi, A.; Vardanyan, T.; Venturi, G. Pauli--Zeldovich cancellation of the vacuum energy divergences, auxiliary fields and supersymmetry. \emph{Eur.\ Phys.\ J.\ C} {\bf 2018}, \emph{78}, 200, doi:10.1140/epjc/s10052-018-5703-6.
\bibitem{Ejlli:2017}
Ejlli, D. Beyond the standard model with sum rules. \emph{arXiv} \textbf{2017}, arXiv:1709.04677 [hep-ph].
  
  
\end{thebibliography}
\end{document}